\documentclass[twocolumn]{aastex63}
\usepackage[version=4]{mhchem}
\usepackage{float}
\usepackage{parskip}
\usepackage{savesym}
\savesymbol{tablenum}
\usepackage{siunitx}
\usepackage{cancel}
\usepackage[normalem]{ulem}
\restoresymbol{SIX}{tablenum}

\DeclareMathOperator\erf{erf}
\received{}
\revised{}
\accepted{}
\shorttitle {Bayesian Inference of Reaction Rate Parameters of a Glycine Network}
\shortauthors{Heyl et al.}
\graphicspath{{./}{figures/}}

\begin{document}

\title{Using Statistical Emulation and Knowledge of Grain-Surface Diffusion for Bayesian Inference of Reaction Rate Parameters: An Application to a Glycine Network}

\correspondingauthor{Johannes Heyl}
\email{johannes.heyl.19@ucl.ac.uk}

\author[0000-0003-0567-8796]{Johannes Heyl}
\affiliation{Department of Physics and Astronomy, University College London, Gower Street, WC1E 6BT, London, UK}

\author[0000-0003-4025-1552]{Jonathan Holdship}
\affiliation{Leiden Observatory, Leiden University, PO Box 9513, 2300 RA Leiden, The Netherlands}
\affiliation{Department of Physics and Astronomy, University College London, Gower Street, WC1E 6BT, London, UK}

\author[0000-0001-8504-8844]{Serena Viti}
\affiliation{Leiden Observatory, Leiden University, PO Box 9513, 2300 RA Leiden, The Netherlands}
\affiliation{Department of Physics and Astronomy, University College London, Gower Street, WC1E 6BT, London, UK}



\begin{abstract}
There exists much uncertainty surrounding interstellar grain-surface chemistry. One of the major reaction mechanisms is grain-surface diffusion for which the the binding energy parameter for each species needs to be known. However, these values vary significantly across the literature which can lead to debate as to whether or not a particular reaction takes place via diffusion. In this work we employ Bayesian inference to use available ice abundances to estimate the reaction rates of the reactions in a chemical network that produces glycine. Using this we estimate the binding energy of a variety of important species in the network, by assuming that the reactions take place via diffusion. We use our understanding of the diffusion mechanism to reduce the dimensionality of the inference problem from 49 to 14, by demonstrating that reactions can be separated into classes. This dimensionality reduction makes the problem computationally feasible. A neural network statistical emulator is used to also help accelerate the Bayesian inference process substantially.

The binding energies of most of the diffusive species of interest are found to match some of the disparate literature values, with the exceptions of atomic and diatomic hydrogen. The discrepancies with these two species are related to limitations of the physical and chemical model. However, the use of a dummy reaction of the form $\ce{H + X -> HX}$ is found to somewhat reduce the discrepancy with the binding energy of atomic hydrogen. Using the inferred binding energies in the full gas-grain version of UCLCHEM results in almost all the molecular abundances being recovered.

\end{abstract}



\section{Introduction} \label{sec:intro}
Interstellar dust plays a very significant role in the rich chemistry that is produced in the interstellar medium. In fact, it is widely believed that complex-organic molecules (COMs) form on interstellar dust \citep{Herbst_Dishoeck, Caselli}. In certain cases, grain-surface reactions are more efficient than gas-phase reactions, due to the dust grains acting as energy sinks. However, there exists much debate about the stage of star formation during which these molecules are produced. While modelling has shown that molecules such as glycine can be formed during the warm-up phase of star formation \citep{Garrod_glycine}, there is also evidence that suggests that dark interstellar cloud conditions would suffice \citep{Ioppolo}. 

Bayesian inference can be used to estimate reaction rate parameters using observations. While this tool has become a staple in many areas of astrophysics, it is only recently that it has found use cases in astrochemistry \citep{Antonios, holdship, Damien, Heyl}. In \cite{holdship}, reaction rates were inferred using a toy network. In \cite{Heyl}, the topology of this network was also considered, specifically the placement of constraints within the network. Both of these works considered the  rates of the reactions, without considering the actual, underlying reaction mechanisms. However, it was noted in both works that the paucity of grain-surface species abundances means that many of the reaction rates will remain undetermined, due to the high levels of degeneracy. This work seeks to circumvent this issue by using the physics of the grain-surface diffusion mechanism to reduce the number of free parameters and therefore break this degeneracy. 

To better understand the importance of various reactions, it is important to have knowledge of the binding energies on dust grains of the species involved. Molecular binding energies provide an upper temperature limit at which the species is still active on the grain surface before it desorbs into the gas phase \citep{Penteado}. As such, having accurate molecular binding energy values is crucial when modelling grain-surface chemistry, as \cite{Penteado} showed that the grain-surface chemistry was very sensitive to the values of the binding energy. A variety of approaches have been taken to determine the binding energies, ranging from experimental approaches \citep{experimental_approach} to density functional theory \citep{Ferrero}. 

However, despite the various approaches used to estimate binding energies, there is still significant uncertainty when it comes to their values. In this work, we use the Bayesian framework to estimate the binding energies of species. This is an important quantity, as it represents the mobility of the species on a dust grain. The values of the binding energies of species differ significantly across the literature \citep{Penteado, UMIST, Wakelam}. This high level of disagreement may be due to differing modelling and/or experimental approaches which cannot necessarily be reconciled. By using measured abundances of some grain-surface species, we are looking to provide estimates of binding energies with uncertainties. 

However, Bayesian inference typically has a long run-time, that is dependent on both the number of dimensions that are being explored as well as the time taken per forward model evaluation. A higher dimensionality means that the Bayesian inference sampler requires more samples to converge to a stationary posterior distribution. We reduce the dimensionality of our problem by utilising physical considerations of the reaction mechanism. This also reduces the total time taken for the inference. We also use statistical emulation to reduce the time further by decreasing the time taken per forward model evaluation. This is particularly relevant when performing the inference multiple times, given that each inference run calls the forward model tens of thousands of times.

We begin by first explaining the chemical code and network that will be used in this work in Section 2. Additionally, we describe the grain-surface diffusion mechanism that lies at the heart of our investigation. We will explain how we can make approximations regarding a species' mobility to estimate the binding energy of said species. In Section 3, we will discuss how statistical emulation can be leveraged to accelerate the running of the forward model, before describing how Bayesian inference will tie all of this together in Section 4. Following this, we will present the resulting binding energies estimates using this method Section 5. In Section 6, we then look to see how well we are able to recover abundances when we run a full gas-grain chemical code using the estimated values.

\section{The Chemical Code and Network}
\subsection{The Chemical Model and Code}
The code that was used was based on the gas-grain chemical code UCLCHEM \citep{UCLCHEM}. 

The surface chemistry is modeled through the rate equation approach. The code has to solve a system of coupled ordinary differential equations of the form:

\begin{equation}\label{rate_equation}
   \frac{dn_{i}}{dt} = \sum_{l,m}k^{i}_{lm}n_{l}n_{m} - n_{i}\sum_{i\neq r}k_{r}n_{r}-k_{i}^{des}n_{i} + k_{i}^{ads}n_{i,gas},
\end{equation}

where $k^{i}_{lm}$is the reaction rate of the reactions between species \textit{l} and \textit{m} to produce species \textit{i}, $n_{i}$ is the abundance of species $i$, $k_{r}$ represents the reaction rates of all reactions where species $i$ is consumed as a reactant and $k_{i}^{des}$ and $k_{i}^{ads}$ represent the desorption and adsorption rates of the species. The coupled differential equations represent the formation and destruction mechanisms for all the relevant species.

However, in order to reduce the runtime of the inference process, some changes had to be made to reduce the time taken for UCLCHEM to run. These are described in detail in \cite{holdship}, but are outlined briefly here. 

The code that was used considered only grain-surface chemistry to reduce the complexity of the system of coupled ordinary differential equations. However, it was important to still include the key processes that couple the gas and grain chemistry. It should be noted that the final two terms in Equation \ref{rate_equation} represent the net flux of gas-phase molecules adsorbing to the grain surface. As such, if one only wishes to consider grain-surface chemistry, then one just needs to parameterize this net ``freeze-out". The net freeze-out was found by running a single point-model of the full gas-grain version UCLCHEM. The net movement of each species between the gas and grain phases as a function of time was then extracted. Only the species which were deposited in abundances relative to $n_{H}$ greater than $10^{-7}$ on the grains were included. These species were: H, O, OH, C, CO, N, CH and CH$_{3}$. These were all species which would form in the gas-phase and were involved in the reactions listed in Table \ref{reaction_network_table}. The freeze-out rates were inserted as source terms into the grain-surface models. The freeze-out of the more complex species was not considered, as these species were unlikely to form in the gas-phase at 10 K. The advantage of doing this is that one avoided needing to consider the system of ODEs for gas-phase reactions, thereby significantly reducing the computational complexity.

The code models the surface chemistry of a collapsing dark cloud from a density of $10^{2}$ cm$^{-3}$ to $10^{6}$ cm$^{-3}$ over 10 million years at 10 K. As in \cite{holdship}, the model reaches its final density at 6 Myr, but the chemistry continues to evolve at constant velocity until the age of the cloud reaches 10 Myr. The grains start off as bare grains, with the freezeout of the gas-phase species acting as source terms for the grain-surface chemistry.

\begin{table}
 \begin{tabular}{||c c c||} 
 \hline
 Reaction No. & Reaction  & $\frac{E_{b}}{E_{D}}$ \\ [0.5ex] 
 \hline
  1 & \ce{H + H -> H_{2}}  & 0.6\\ 
 \hline
 2 & \ce{O + H -> OH} & 0.6 \\ 
 \hline
 3 & \ce{OH + H -> H_{2}O} & 0.6  \\
 \hline
 4 & \ce{CO + H -> HCO} & 0.6  \\  
 \hline
 5 & \ce{HCO + H -> H_{2}CO}  & 0.6 \\  
 \hline
 6 & \ce{HCO + H -> H_{2} + CO}  & 0.6 \\  
 \hline
 7 & \ce{H_{2}CO + H -> H_{3}CO} & 0.6  \\  
 \hline
 8 & \ce{H_{2}CO + H -> HCO + H_{2}} & 0.6\\
 \hline
 9 & \ce{H_{3}CO + H -> CH_{3}OH} & 0.6  \\
 \hline
 10 & \ce{CO + OH -> HOCO} & 0.5\\
 \hline
 11 & \ce{CO + OH -> CO_{2}} & 0.5\\
 \hline
 12 & \ce{HOCO + H -> H_{2} + CO_{2}} & 0.6\\
 \hline
 13 & \ce{HOCO + H -> HCOOH} & 0.6\\
 \hline
 14 & \ce{N + H -> NH} & 0.6\\
 \hline
 15 & \ce{NH + H -> NH_{2}} & 0.6\\
 \hline
 16 & \ce{NH_{2} + H -> NH_{3}} & 0.6\\
 \hline
 17 & \ce{C + H -> CH} & 0.6\\
 \hline
 18 & \ce{CH + H -> CH_{2}} & 0.6\\
 \hline
 19 & \ce{CH_{2} + H -> CH_{3}} & 0.6\\
 \hline
 20 & \ce{CH_{3} + H -> CH_{4}} & 0.6\\
 \hline
 21 & \ce{CH_{4} + OH -> CH_{3} + H_{2}O} & 0.6\\
 \hline
 22 & \ce{NH_{2} + CH_{3} -> NH_{2}CH_{3}} & 0.5\\
 \hline
 23 & \ce{NH_{3} + CH -> NCH_{4}} & 0.5\\
 \hline
 24 & \ce{NCH_{4} + H -> NH_{2}CH_{3}} & 0.6\\
 \hline
 25 & \ce{NH_{2}CH_{3} + H -> NCH_{4} + H_{2}} & 0.6\\
 \hline
 26 & \ce{NH_{2}CH_{3} + OH -> NCH_{4} + H_{2}O} & 0.5\\
 \hline
 27 & \ce{NCH_{4} + HOCO -> NH_{2}CH_{2}COOH} & 0.5\\
 \hline
 28 & \ce{OH + H_{2} -> H_{2}O} & 0.35\\
 \hline
 29 & \ce{O + O -> O_{2}} & 0.6\\
 \hline
 30 & \ce{O_{2} + H -> HO_{2}} & 0.6\\
 \hline
 31 & \ce{HO_{2} + H -> OH + OH} & 0.6\\
 \hline
 32 & \ce{HO_{2} + H -> H_{2} + O_{2}} & 0.6\\
 \hline
 33 & \ce{HO_{2} + H -> H_{2}O + O} & 0.6\\
 \hline
 34 & \ce{OH + OH -> H_{2}O_{2}} & N/A\\
 \hline
 35 & \ce{OH + OH -> H_{2}O + O} & N/A\\
 \hline
 36 & \ce{H_{2}O_{2} + H -> H_{2}O + OH} & 0.6\\
 \hline
 37 & \ce{N + N -> N_{2}} & 0.6\\
 \hline
 38 & \ce{N + O -> NO} & 0.6\\
 \hline
 39 & \ce{NO + H -> HNO} & 0.6\\
 \hline
 40 & \ce{HNO + H -> H_{2}NO} & 0.6\\
 \hline
 41 & \ce{HNO + H -> NO + H_{2}} & 0.6\\
 \hline
 42 & \ce{HNO + O -> NO + OH} & 0.6\\
 \hline
 43 & \ce{HN + O -> HNO} & 0.6\\
 \hline
 44 & \ce{N + NH -> N_{2}} & 0.6\\
 \hline
 45 & \ce{NH + NH -> N_{2} + H_{2}} & 0.5\\
 \hline
 46 & \ce{C + O -> CO} & 0.6\\
 \hline
 47 & \ce{CH_{3} + OH -> CH_{3}OH} & 0.6\\
 \hline
 48 & \ce{NH + CO -> HNCO} & 0.5\\
 \hline
 49 & \ce{NH_{3} + HNCO -> NH_{4}^{+} + OCN^{-}} & 0.35\\
 \hline
\end{tabular}
\caption{Table of the reactions used in this work taken from \cite{Ioppolo} and \cite{Linnartz_review}. The values of $\frac{E_{b}}{E_{D}}$ used for the more mobile species for each diffusion-based reaction are given. Reactions 34 and 35 are not assumed to be diffusive reactions.}
\label{reaction_network_table}
\end{table}

\subsection{The Chemical Network}
Our network is composed of radicals that react to form glycine, the simplest amino acid. The reactions that make up this chemical network are shown in Table \ref{reaction_network_table}. The grain-surface network used in this work is based on the one used in \cite{Ioppolo} with the final two reactions being taken from \cite{Linnartz_review}. In \cite{Ioppolo}, laboratory and chemical modelling found that the first 47 reactions were able to produce glycine in dark interstellar conditions, long before the warm-up phase of star formation, without requiring any energetic input \citep{Ioppolo}. This is in contrast to previous work that assumed that the formation of glycine required an increased temperature as well as energetic processing \citep{Garrod_glycine}. Based on \cite{Ioppolo}, it was expected that this network would be sufficient to learn about COMs in the pre-stellar phase with the help of observed abundances. Reactions 48 and 49 were included, as they involved species already present in the network. Furthermore, one of the end-products, NH$_{4}^{+}$, had a constraint on its abundance that could be used for the Bayesian inference to further constrain the parameters.

\subsection{Grain Surface Chemistry}
\subsubsection{Grain Surface Diffusion}
An understanding of the actual grain surface mechanisms will prove crucial in this work. The diffusion mechanism described in \cite{Hasegawa} was implemented in UCLCHEM in \cite{Quenard}.

According to the mechanism, the rate at which two species A and B react via diffusion is given by: 

\begin{equation}\label{reaction_rate}
k_{AB} = \kappa_{AB}\frac{(k^{A}_{hop}+k^{B}_{hop})}{N_{site}n_{dust}},
\end{equation}

where $N_{site}$ is the number of sites on the grain surface and $n_{dust}$ is the number density of dust grains.




In equation \ref{reaction_rate}, $k^{X}_{hop}$ is the thermal hopping rate of species $X$ on the grain surface defined as: 

\begin{equation}\label{hopping_rate_equation}
k^{X}_{hop} = \nu_{0}\exp\left(-\frac{E_{b}}{T_{gr}}\right),
\end{equation}

where $E_{b}$ is the diffusion energy of the species, $T_{gr}$ is the grain temperature and $\nu_{0}$ is the characteristic vibration frequency of species $X$. The diffusion energy is typically taken to be a fraction of the species binding energy, $E_{D}$. There is debate surrounding the value of the fraction $\frac{E_{b}}{E_{D}}$, though there is agreement that it should be in the range 0.3 to 0.8, with lower values in this range being more appropriate for stable molecules \citep{Penteado}. However, it has been found that for O and N atoms, a ratio of 0.55 is more suitable  \cite{diffusion_binding_ratio}. In this work, we follow the convention adopted by \cite{non_diffusive_mechanisms} where the ratio was set to equal 0.6 for atomic species and 0.35 for stable species. For all other species, a value of 0.5 was used. For each reaction, the value of $\frac{E_{b}}{E_{D}}$ for the more mobile species is given in Table \ref{reaction_network_table}. The reason we only consider the value of this ratio for the more mobile species is given in Section \ref{binding_energy_derivation}. The characteristic vibration frequency, $\nu_{0}$, is defined as: 

\begin{equation}
\nu_{0} = \sqrt{\frac{2k_{b}n_{s}E_{D}}{\pi^{2}m}},
\end{equation}

where $k_{b}$ is the Boltzmann constant, $n_{s}$ is the grain site density and $m$ is the mass of species $X$. 




Finally, $\kappa_{AB}$, which provides the reaction probability is taken to be:

\begin{equation}
\kappa_{AB} = \max\left(\exp{\left(-\frac{2a}{\hbar}\sqrt{2\mu k_{b}E_{A}}\right)}, \exp{\left(-\frac{E_{A}}{T_{gr}}\right)}\right)
\end{equation}, 

where $\hbar$ is the reduced Planck constant, $\mu$ is the reduced mass, $E_{A}$ is the reaction activation energy, $k_{b}$ is Boltzmann's constant and $a = 1.4 \si{\angstrom}$ is the thickness of a quantum mechanical barrier. The reaction probability is effectively a competition between the first term, which is the quantum mechanical probability of a tunnelling through a rectangular barrier of thickness $a$ and the thermal reaction probability, the second term.

\subsubsection{Reaction-diffusion competition}
A correction needs to be made to $\kappa_{AB}$ to account for the fact that species might diffuse or evaporate instead of reacting with each other. This correction is the reaction-diffusion competition \citep{Chang, GarrodandPauly}. The reaction probability is defined to be: 

\begin{equation}
\kappa_{AB}^{final} = \frac{p_{reac}}{p_{reac} + p_{diff} + p_{evap}},
\end{equation}

where $p_{reac}$, $p_{diff}$ and $p_{evap}$ represent the probabilities of species A and B reacting, diffusing and evaporating per unit time, respectively. These quantities are defined as: 

\begin{equation}
p_{reac} = \max(\nu_{0}^{A}, \nu_{0}^{B})\kappa_{AB}
\end{equation}, 

\begin{equation}
p_{diff} = k_{hop}^{A} + k_{hop}^{B}
\end{equation}

and

\begin{equation}
p_{evap} = \nu_{0}^{A}\exp\left(-\frac{E_{D}^{A}}{T_{gr}}\right) + \nu_{0}^{B}\exp\left(-\frac{E_{D}^{B}}{T_{gr}}\right).
\end{equation}

In equation \ref{reaction_rate}, $\kappa_{AB}$ is replaced with $\kappa_{AB}^{final}$.

At 10K, we observe that the rate of evaporation is far lower than the rates of diffusion and reaction, so will be neglected throughout this work.  

As most of the reactions in Table \ref{reaction_network_table} are radical-radical, it was assumed that their activation energies were 0K. Even for reactions involving a known reaction barrier, such as reaction 5, it was found that $p_{reac} \gg p_{diff}$, which means the activation energy barrier is lower than the diffusion barrier. As such, $\kappa_{AB}^{final} \simeq 1$. This ``diffusion-limited regime" corresponds to the situation where the diffusion process is the rate-limiting step and is due to the fact that the temperature being considered is 10K.


\section{Statistical Emulation}
Statistical emulation involves fitting a statistical function to match the inputs and outputs of a forward model \citep{emulator_definition}. The advantage in doing so is that one replace the slow-to-evaluate forward model with the fitted emulator in order to save time. This becomes particularly significant when multiple evaluations of the forward model are required, such as in Bayesian inference which typically involves calling the forward model hundreds of thousands of times. Statistical emulators have primarily been used in the past in cosmology \citep{Auld, Wang, Rogers_2019, Pritchard}, but have also recently found use in astrochemistry \citep{Damien, chemulator}. In our case, the forward model requires solving a coupled system of ODEs of the form given in Equation \ref{rate_equation}. The evaluation of the forward model can be time-consuming, especially if this has to be repeated multiple times as would be the case for Bayesian inference. This proves to be particularly important for the analysis we do in Appendices \ref{frequentist_properties_bayesian_estimator} and \ref{binding_energy_as_function_of_constraint}. In this case, a statistical emulator would be particularly useful, as it can interpolate within the range of input values considered. This is computationally faster than making use of the original forward model for evaluation.

There are a number of algorithms that can be used for the purposes of emulation. One particularly popular one is the Gaussian process emulator, which has found widespread usage \citep{kennedyohagan, gaussian_processes_cosmological_simulations, Rogers_2019}. An inherent advantage is the ability of this sort of emulator to quantify the uncertainty associated with the regression. This allows for the use of acquisition function that iteratively improve the emulator approximation by sampling points in areas of high uncertainty \citep{gaussian_processes_cosmological_simulations, Rogers_2019}. However, a disadvantage is that the emulation process scales badly as the cube of number of training points 
\citep{gaussian_processes_cosmological_simulations}. This is in contrast to neural network emulators, which will be used in this work. Neural networks aim to fit the relationship between the inputs and outputs of the model without considering the uncertainty of the approximation. Neural networks do not struggle as drastically with an increase in training points. A higher number of training points will ensure better model performance as the emulator, which is the reason that we elected to use neural networks.

\begin{figure*}
    \includegraphics{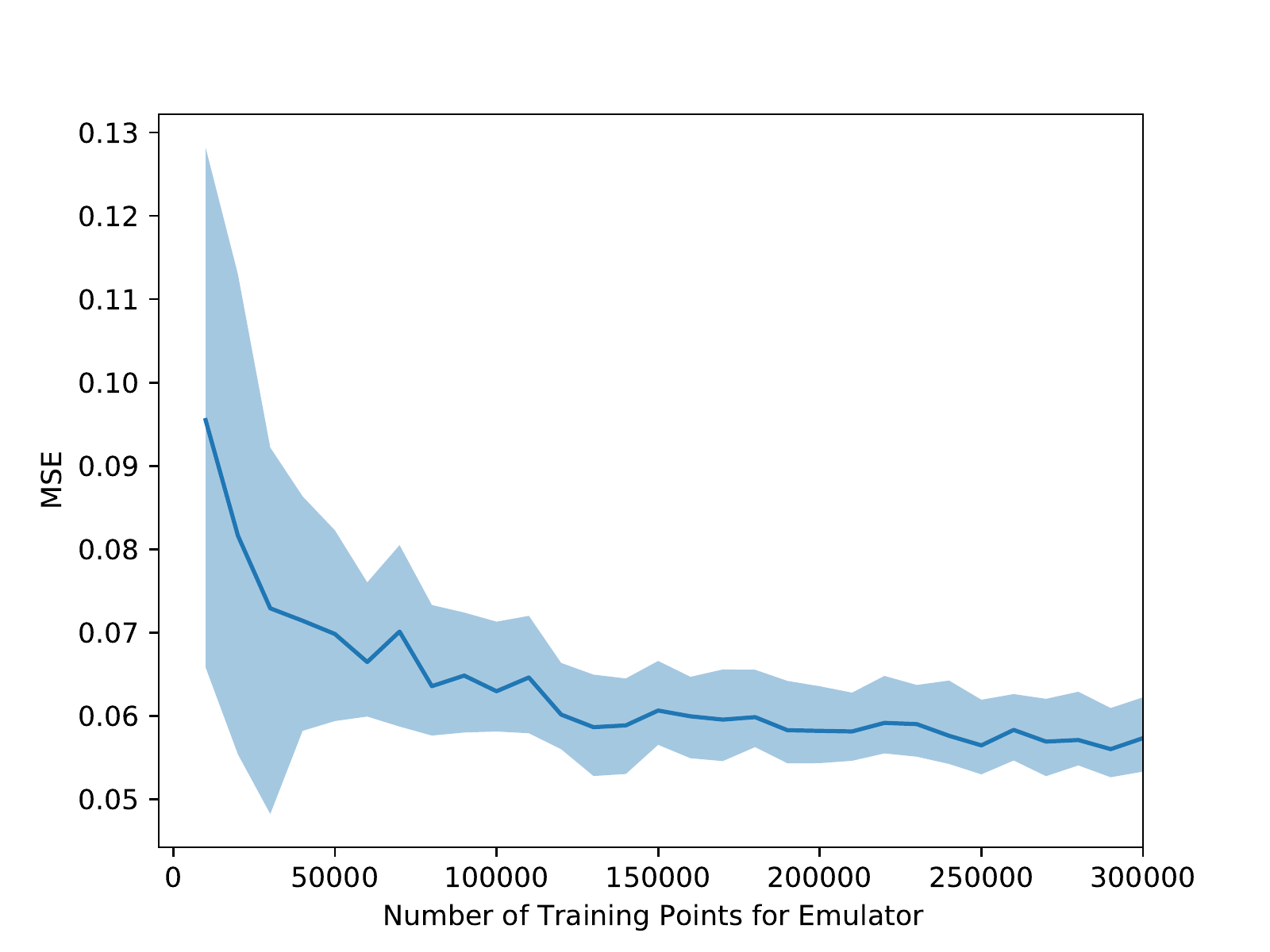}
    \caption{A plot of the mean-squared error of the emulator as a function of the number of training points used to train the emulator. The shaded area represents the 95\% confidence interval around the mean-squared error.}
    \label{mse_vs_training_points}
\end{figure*}

\subsection{Training the Emulator}
In order to be able to use the emulator, it must first be trained on some data. It is important that the sampling is done in such a way that the entire parameter space is explored. One cannot simply use random uniform sampling, as each point is drawn independently of the others. This can result in the training points being clustered. This has the consequence of the emulator attempting to match the training data more in these regions, thereby introducing bias in other less well-covered regions of the parameter space. A Latin Hypercube Sampling Scheme was used \citep{LHS} and implemented using the Python surrogate modelling toolbox \citep{LHS_python}. As both the input and output parameters span several orders of magnitude, the emulator was trained to learn the mapping between the logarithm of these two. The training dataset spanned the prior range for each parameter. Given that a log-uniform prior between $10^{-15}$ and $10^{0}$ was used for the Bayesian inference (see Section \ref{implementation} for details), this ensured that any conceivable input to the emulator from the inference was within the prior range, as outside that range the posterior is zero due to the prior being zero. The parameter ranges defined the range of values over which the emulator could interpolate. The emulator was not needed to extrapolate, as the range of the prior was covered.

Choosing the number of training points is a crucial parameter. It is clear that increasing the amount of training data will improve the emulator performance. However, this will also result in the time taken for training increasing. As such, a balance needs to be struck. Figure \ref{mse_vs_training_points} shows the mean-squared error (MSE) on a test set as a function of the number of training points. It was found that using 150,000 training points was sufficient. By evaluating these points on a single RCIF node with 40 cores, the training time was about 30 minutes.

\subsection{The Neural Network}
In this work, an artificial neural network was used as the emulator. To improve the neural network's performance, the input log-rates were scaled to lie between zero and one. A five-layer neural network was used with the three hidden layers containing 512, 256 and 128 neurons, respectively. The hyperbolic tangent was used as the activation function. The scikit-learn package was used to train the emulator \citep{scikit-learn}. To avoid over-fitting to the training data, the training process was terminated when the validation error stopped decreasing by at least 0.01.

\section{Bayesian Inference}

\subsection{Introduction to Bayesian Inference}

The aim of this work is to deduce the reaction rates of the reactions in this network, which we represent as a vector, $\textbf{k} = (k_{1}, k_{2} ... k_{49})$, and use these inferred reaction rates to determine the binding energies of diffusive species. This is initially a 49-dimensional inference problem. The code used takes this vector as an input and outputs the abundances of all the species in this network, which is represented by the vector $\textbf{Y} = (Y_{1}, Y_{2} ... Y_{35})$. There exist measurements for the abundances of a subset of the molecules in this network. These form the data \textbf{d}, which are listed in Table \ref{abundance_table}. 

Bayes' Law can be used to determine the probability distribution of the reaction rates given the data: 

\begin{equation}
P(\textbf{k} \vert \textbf{d}) = \frac{P(\textbf{d} \vert \textbf{k})P(\textbf{k})}{P(\textbf{d})},
\end{equation}
where $P(\textbf{k} \vert \textbf{d})$ is the posterior probability distribution, $P(\textbf{k})$ is the prior, $P(\textbf{d} \vert \textbf{k})$ is the likelihood and $P(\textbf{d})$ is referred to as the evidence. The prior distribution encodes our initial knowledge of the values of the reaction rates. The likelihood provides the likelihood of the data as a function of the reaction rates. The likelihood provides information about the physical model under consideration. The evidence serves as a normalising factor, as it represents the marginalised likelihood. The posterior distribution represents the updated probability distribution of reaction rates given the data, encoded in the prior distribution, and the physical model.

\subsection{Implementation}\label{implementation}
To obtain the posteriors of the reaction rates, a prior must be specified. As has been done previously, a log-uniform prior was chosen, so as to equally weight rates over different orders of magnitude. However, a different range is chosen compared to \cite{holdship} and \cite{Heyl}, to accommodate the fact that the reaction rates, \textbf{k}, are normalised by the cloud density. Additionally, it was found in \cite{holdship} and \cite{Heyl} that the probability density is very low in the range $10^{-30} - 10^{-15}$. As such, a log-uniform prior between $10^{-15}$ and $10^{0}$ was used.

We assume that the measurements are Gaussian based on the fact the distribution of reported measurements such as in \cite{Whittet} are not strongly skewed  but instead are reasonably well fit by Gaussians with the parameters we include in our data table. A Gaussian likelihood function was used: 

\begin{equation}\label{likelihood}
\centering
P(\textbf{d} \vert \textbf{k}) =  \prod_{i=1}^{n_{d}} \frac{1}{\sqrt{2\pi}\sigma_{i}} \exp\left({-\frac{(d_{i}-Y_{i})^{2}}{2\sigma_{i}^{2}}}\right),
\end{equation}

where $n_{d}$ is the number of observations and $\sigma_{i}$ is the uncertainty of the $i$th observation. Only the species for which there are abundances are multiplied over. Table \ref{abundance_table} contains species for which we have abundances with Gaussian uncertainties. Observed abundances will be referred to as constraints in this work as they constrain the prior parameter space of reaction rate posteriors.

\cite{Boogert} also contains upper limits for the abundances of some species of interest. The upper limits for O$_2$, N$_2$, H$_2$O$_2$ and glycine are also included in Table \ref{abundance_table}. Equation \ref{likelihood} can be rewritten to account for these upper limits, as was done in \cite{holdship}.

\begin{equation}\label{upper_limit_likelihood}
\centering
P(\textbf{d} \vert \textbf{k}) =  \prod_{i=1}^{n_{d}} \frac{1}{\sqrt{2\pi}\sigma_{i}} \exp\left({-\delta_{i}\frac{(d_{i}-Y_{i})^{2}}{2\sigma_{i}^{2}}}\right)(1-S(C_{i}))^{1-\delta_{i}},
\end{equation}

where $\delta_{i}$ is 1 for observed species and 0 for species with upper limits. Notice that in this case that $n_{d}$ is the number of observations as well as upper limits. $C_{i}$ is the upper limit of that species and $S(C_{i})$ is the survival function, which is defined as

\begin{equation}\label{survival_function}
\centering
S(C_{i}) = 1 - \frac{1}{2}\left(1 + \erf\left(\frac{C_{i}-Y_{i}}{\sigma_{i}^{UL}}\right)\right),
\end{equation}

where $\erf$ is the error function and \textbf{$\sigma_{i}^{UL}$} is taken to be one-third of the upper limit. The value of $\sigma_{i}$ is to account for the fact that there might be some level of uncertainty on the value of the upper limit.

In order to sample the posterior, the PyMultiNest Python package was used \citep{pymultinest}, which is a wrapper for the MultiNest package \citep{MultiNest1, MultiNest2, MultiNest3}, which implements nested sampling \citep{Skilling}. A Python wrapper of the UCLCHEM code was created using F2Py. The input was the vector of reaction rates \textbf{k}.

\subsection{Degeneracy Problem}
Before performing the inference, it is important to consider the problem in more depth. There are 49 parameters to estimate, but there are only 12 measurements. As was observed in \cite{holdship} and \cite{Heyl}, having far more parameters than constraints introduces a significant amount of degeneracy into the problem. Some rates of reactions do not influence the abundances of species with constraints. As was observed in \cite{holdship}, this will result in the majority of reaction rate posteriors being uniform. Additionally, many posteriors will only deviate weakly from uniformity. This was found to be the case for linear reaction chains with successive hydrogenations, such as the successive hydrogenation of CO to form methanol. The degeneracy stemmed from the fact that the reactions were tightly coupled. Provided one rate took a minimum value and acted as the rate-limiting step, the other reaction rate was free to vary above this minimum rate. The high level of degeneracy inherent to this problem meant that despite running a sampler for several weeks, it never converged. 

\subsection{Degeneracy Solution}

To reduce the degeneracy of the problem, one can exploit information about the underlying grain-surface diffusion mechanism. Ultimately, the reaction rate is strongly dependent on the hopping rates of the reactant species, which, assuming the grain temperature is constant, implies that the reaction rate is set by the binding energies. Given the strong dependence of the hopping rate on the binding energy, it is clear that a small difference in the binding energy between two species will mean that the  hopping rate of the more mobile species (the one with the lower binding energy) will dominate the reaction rate. In equation \ref{reaction_rate}, this corresponds to  $k^{A}_{hop} \gg k^{B}_{hop}$ and yields 

\begin{equation}\label{reaction_rate_modified}
k_{AB} = \kappa_{AB}^{final}\frac{k^{A}_{hop}}{N_{site}n_{dust}},
\end{equation}

where we see that this equation only depends on the hopping rate of species $A$. Recall that $\kappa_{AB}^{final} \simeq 1$ in the diffusion-limited regime.

Based on this, one can separate reactions into various classes, depending on which of their reactants is more mobile. Even though the actual values of the binding energies will differ across the literature (see \cite{Penteado} for a discussion on this), most works agree on the ``hierarchy" of mobility, that is which of the two species is more mobile. By making an assumption or by considering literature values, one can make a decision on which species should be treated as more mobile. In this work, the more mobile species was assumed to be the one with the lower binding energy in at least two of \cite{Penteado}, UMIST and \cite{Wakelam}. The groupings used are shown in Table \ref{table_of_groupings}. For reactions 34 and 35, one does not expect diffusion to be the dominant reaction mechanism. However, since this is the diffusion-limited regime, one can assume that these two reactions will have the same reaction rate. 

The major implication is that this now allows for the calculation of a species' binding energy. In fact, provided that species is far more mobile, one can calculate that species' binding energy. What one finds is that the reaction rates of many reactions are effectively only dependent on the binding energy of the same species. As such, the dimensionality of the problem is significantly reduced, as one simply needs to determine the binding energies of the more mobile species.

\subsection{Deriving the Binding Energies}\label{binding_energy_derivation}
Binding energy values vary greatly across the literature. Their values can determine whether or not a reaction can occur efficiently via diffusion. For example in \cite{Ioppolo}, it is stated that 10K is too low a temperature for any species other than atomic hydrogen to diffuse. However, a statement such as this one assumes a value for the binding energy of hydrogen and that it is far lower than the binding energies for other species. While many works state the binding energy of H to be 650K, many others find that species such as O and N have comparable binding energies \citep{Penteado}.

Our goal is to determine the binding energies of various species. In this work, we will be inferring reaction rates for the various reactions and use these to solve for the binding energies. The quantity \textbf{k} varies as a function of time, as seen in equation \ref{reaction_rate} due to the dependence on the total hydrogen number density. However, by multiplying by $n_{H}$ on both sides, one obtains

\begin{equation}
k'_{AB} = k_{AB}n_{H} = \kappa_{AB}\frac{(k^{A}_{hop}+k^{B}_{hop})}{N_{site}\frac{n_{dust}}{n_{H}}},
\end{equation}

where $\frac{n_{dust}}{n_{H}}$ is a constant. Note now that the expression on the right-hand side only consists of constants. This implies that $k'_{AB}$ is constant with respect to the density of the cloud. While $k'_{AB}$ can still be interpreted as a reaction rate, it has units of $s^{-1}$.

Due to the exponential dependence of the hopping rates on the binding energies, one finds that in most cases, one species dominates the reaction rate. If species A has a binding energy of, say, 500K and species B has a 10\% higher binding energy, then A's hopping rate is almost 150 times greater, due to the low grain temperature of 10K. This difference will only get larger as the binding energies under consideration increase. Hence, we can state that $k^{A}_{hop} \gg k^{B}_{hop}$ and then determine the binding energy of the species by substituting equation \ref{hopping_rate_equation} into equation \ref{reaction_rate_modified}:

\begin{equation}\label{equation_to_solve_for_E_D}
k'_{AB}\frac{n_{dust}N_{site}}{n_{H}\kappa_{AB}} \sqrt{\frac{\pi^{2}m}{2k_{b}n_{s}}}= \sqrt{\frac{E^{A}_{b}}{f}}\exp\left(-\frac{E^{A}_{b}}{T_{gr}}\right),
\end{equation}

where the corresponding value of $f  = \frac{E_{b}}{E_{D}}$ is used, depending on the species under consideration. This equation cannot be solved analytically, so has to be solved numerically.

\subsection{Constraints}
The final component required to perform Bayesian inference is the data, which in this case would be measured abundances of species. A number of constraints for molecules in this network can be found in \cite{Boogert}, which provides the median abundance as well as lower and upper quartile. As in \cite{holdship}, we assume the measurements are Gaussian-distributed, which implies the median is the mean. Additionally, the upper and lower quartiles are $0.68\sigma$ from the mean. Using this information, the abundances used in this work are listed in Table \ref{abundance_table}. We combine measured molecular ice abundances from dark, quiescent cloud as well as Large Young Stellar Objects (LYSOs). We observe that the species CO, CO$_{2}$, H$_{2}$O, CH$_{3}$OH and NH$_{4}^{+}$ have similar abundances in quiescent clouds and LYSOs. Using this, we assume that other species, which have only been detected in LYSOs, will have broadly similar dark cloud abundances. We argue that while chemistry is expected to happen during the warm-up phase for LYSOs, this will be relatively short-lived and any abundances will likely have been built up during the cold phase of star formation. However, even though the warm-up phase will be shorter, the chemical time scales will decrease due to the reaction rate's dependence on temperature. Overall, while there is justification for using LYSO abundances for dark cloud conditions, it should be noted that we are adding additional uncertainty into our analysis.


\section{Results}
\begin{table}
 \begin{tabular}{||c c c||} 
 \hline
 Species & Abundances relative to H  & Source\\ [1ex] 
 \hline\hline
 H$_{2}$O & $(4.0 \pm 1.3) \times 10^{-5}$ &  Cloud \\ 
 \hline
 CO & $(1.2 \pm 0.8) \times 10^{-5}$ &  Cloud\\
 \hline
 CO$_{2}$ & $(1.3 \pm 0.7) \times 10^{-5}$ &  Cloud\\
 \hline
 CH$_{3}$OH & $(5.2 \pm 2.4) \times 10^{-6}$ &  Cloud\\
 \hline
  NH$_{3}$ & $(3.6 \pm 2.6) \times 10^{-6}$ & LYSOs\\
 \hline
 CH$_{4}$ & $(2.3 \pm 2.1) \times 10^{-6}$ & LYSOs\\
 \hline
  HCOOH & $(2.4 \pm 1.3) \times 10^{-6}$ & LYSOs \\
 \hline
  NH$_{4}^{+}$ & $(3.8 \pm 1.5) \times 10^{-6}$ &  Cloud\\
 \hline
  O$_{2}$ & $<60\times 10^{-6}$ & Comet\\
 \hline
  N$_{2}$ & $<0.1-28 \times 10^{-6}$ & Comet\\
 \hline
  H$_{2}$O$_{2}$ & $<0.6-8 \times 10^{-6}$ & Comet\\
 \hline
 NH$_{2}$CH$_{2}$COOH & $<0.1 \times 10^{-6}$ & Comet\\

 \hline
\end{tabular}
\caption{The abundances and uncertainties taken for the network adapted from \cite{Boogert}. There were two distinct values for the upper limit on the abundance of O$_{2}$, so the higher one was selected.}
\label{abundance_table}
\end{table}

\begin{table}[]
\begin{tabular}{|l|l|}
\hline
\textbf{Grouping}   & \textbf{Reactions in Group}          \\ \hline
Hydrogenations      & 1-9, 12-20, 24, 25, 30-33, 36, 39-41 \\ \hline
Oxygenations        & 29, 42, 43                           \\ \hline
Nitrogenations      & 37, 38, 44                           \\ \hline
CO-based reactions              & 10, 11, 48                               \\ \hline
OH+OH               & 34, 35                               \\ \hline
CH3-based reactions & 22, 47                               \\ \hline
\end{tabular}
\caption{The main reaction groupings, separated by the molecule that the literature suggested was more dominant. Any reaction not included in this table had its reaction rate inferred separately.}
\label{table_of_groupings}
\end{table}

\hspace*{-10cm}\begin{table*}[!htbp]
\begin{tabular}{llllll}
\hline
\multicolumn{1}{|l|}{\textbf{Species}} &
  \multicolumn{1}{l|}{\textbf{Binding Energy 1 (K)}} &
   \multicolumn{1}{l|}{\textbf{Binding Energy 2 (K)}} &
  \multicolumn{1}{l|}{\textbf{Penteado (K)}} &
  \multicolumn{1}{l|}{\textbf{Wakelam (K)}} &
  \multicolumn{1}{l|}{\textbf{UMIST (K)}} \\ \hline
\multicolumn{1}{|l|}{H}    & \multicolumn{1}{l|}{$1099^{+33}_{-57}$} & \multicolumn{1}{l|}{$1016^{+65}_{-68}$}   & \multicolumn{1}{l|}{$650\pm 100$}  & \multicolumn{1}{l|}{650}  & \multicolumn{1}{l|}{600}  \\ \hline
\multicolumn{1}{|l|}{O}    & \multicolumn{1}{l|}{$824^{+180}_{-109}$}   & \multicolumn{1}{l|}{$805^{+88}_{-97}$}& \multicolumn{1}{l|}{$1660 \pm 60$}  & \multicolumn{1}{l|}{1600} & \multicolumn{1}{l|}{800}  \\ \hline
\multicolumn{1}{|l|}{N}    & \multicolumn{1}{l|}{$894^{+326}_{-202}$}   & \multicolumn{1}{l|}{$932^{+102}_{-130}$} & \multicolumn{1}{l|}{$715 \pm 358$}  & \multicolumn{1}{l|}{720}  & \multicolumn{1}{l|}{800}  \\ \hline
\multicolumn{1}{|l|}{C}    & \multicolumn{1}{l|}{$1336^{+136}_{-160}$}   &
\multicolumn{1}{l|}{$1361^{+124}_{-256}$}   &
\multicolumn{1}{l|}{$715 \pm 360$}  & \multicolumn{1}{l|}{10000}  & \multicolumn{1}{l|}{800}  \\ \hline
\multicolumn{1}{|l|}{CO}   & \multicolumn{1}{l|}{$1009^{+158}_{-123}$} &
\multicolumn{1}{l|}{$1018^{+91}_{-135}$} &\multicolumn{1}{l|}{$1100 \pm 250$}  & \multicolumn{1}{l|}{1300} & \multicolumn{1}{l|}{1150} \\ \hline
\multicolumn{1}{|l|}{CH}   & \multicolumn{1}{l|}{$1160^{+130}_{-240}$} & \multicolumn{1}{l|}{$1107^{+228}_{-162}$}  & \multicolumn{1}{l|}{$590 \pm 295$}  & \multicolumn{1}{l|}{925}  & \multicolumn{1}{l|}{925}  \\ \hline
\multicolumn{1}{|l|}{CH$_{3}$}    & \multicolumn{1}{l|}{$1088^{+375}_{-242}$}  &
\multicolumn{1}{l|}{$1133^{+350}_{-288}$}  &\multicolumn{1}{l|}{$1040 \pm 500$}  & \multicolumn{1}{l|}{1600}  & \multicolumn{1}{l|}{1175}  \\ \hline
\multicolumn{1}{|l|}{CH$_{4}$}    & \multicolumn{1}{l|}{$1343^{+765}_{-200}$}  &
\multicolumn{1}{l|}{$1327^{+153}_{-139}$}  &
\multicolumn{1}{l|}{$1250 \pm 120$}  & \multicolumn{1}{l|}{960}  & \multicolumn{1}{l|}{1090}  \\ \hline
\multicolumn{1}{|l|}{H$_{2}$}    & \multicolumn{1}{l|}{$1719^{+377}_{-356}$}  &
\multicolumn{1}{l|}{$1976^{+184}_{-283}$}  &\multicolumn{1}{l|}{$500 \pm 100$}  & \multicolumn{1}{l|}{440}  & \multicolumn{1}{l|}{430}  \\ \hline
\multicolumn{1}{|l|}{NH}    & \multicolumn{1}{l|}{$1172^{+307}_{-325}$}  &
\multicolumn{1}{l|}{$1115^{+351}_{-254}$}  &
\multicolumn{1}{l|}{$542 \pm 270$}  & \multicolumn{1}{l|}{2600}  & \multicolumn{1}{l|}{2378}  \\ \hline
\multicolumn{1}{|l|}{NCH$_{4}$} & \multicolumn{1}{l|}{$1265^{+206}_{-354}$}  &
\multicolumn{1}{l|}{$1046^{+326}_{-222}$}  &
\multicolumn{1}{l|}{-} & \multicolumn{1}{l|}{-}    & \multicolumn{1}{l|}{-}    \\ \hline
\multicolumn{1}{|l|}{NH$_{2}$CH$_{3}$} & \multicolumn{1}{l|}{$1694^{+486}_{-355}$}  &
\multicolumn{1}{l|}{$1581^{+544}_{-427}$}  & \multicolumn{1}{l|}{-} & \multicolumn{1}{l|}{-}    & \multicolumn{1}{l|}{-}    \\ \hline
\end{tabular}
\caption{The binding energies obtained for various species obtained through the use of Bayesian inference as well as values from \cite{Penteado}, \cite{UMIST} and \cite{Wakelam}. The first set of predicted binding energies come from performing Bayesian inference on the standard network, while the second set of predictions stem from including the dummy reaction \ce{H + X -> HX}. With the exception of H, most of the other binding values match at least one literature value. For most of the species, the uncertainty on the binding energy values is lower compared to the spread of literature values.  No values for the binding energies of NCH$_{4}$ and NH$_{2}$CH$_{3}$ were found in the literature.}
\label{binding_energy_table}
\end{table*}

\subsection{Highest Density Regions}
Parameter estimates are typically quoted by considering the marginalised posterior distributions. The important quantities to estimate are typically the mean and variance. However, one must be careful when estimating these quantities, as depending on how broad and asymmetric the posterior space is around the maximum-posterior value, these might not be meaningful quantities. To determine useful estimators, one can choose to only consider the highest density region (HDR) of the posterior. 

For a probability density function $f(x)$ for some random variable $X$, the $100(1-a)\%$ HDR is the subset $R(f_{a})$ of values in $X$ such that 

\begin{equation}
R(f_{a}) = x : f(x) \geq f_{a}, 
\end{equation}

where $f_{a}$ is the largest constant that ensures that the probability of being in $R(f_{a})$ is greater than 1-a \citep{Hyndman}. In other words, the HDR allows one to only consider a subset of the posterior density function that has a value greater than some threshold $f_{a}$. 

\subsection{Reaction Rate Marginalised Posteriors}

\begin{figure*}[h]
    \centering
    \includegraphics[width=\textwidth]{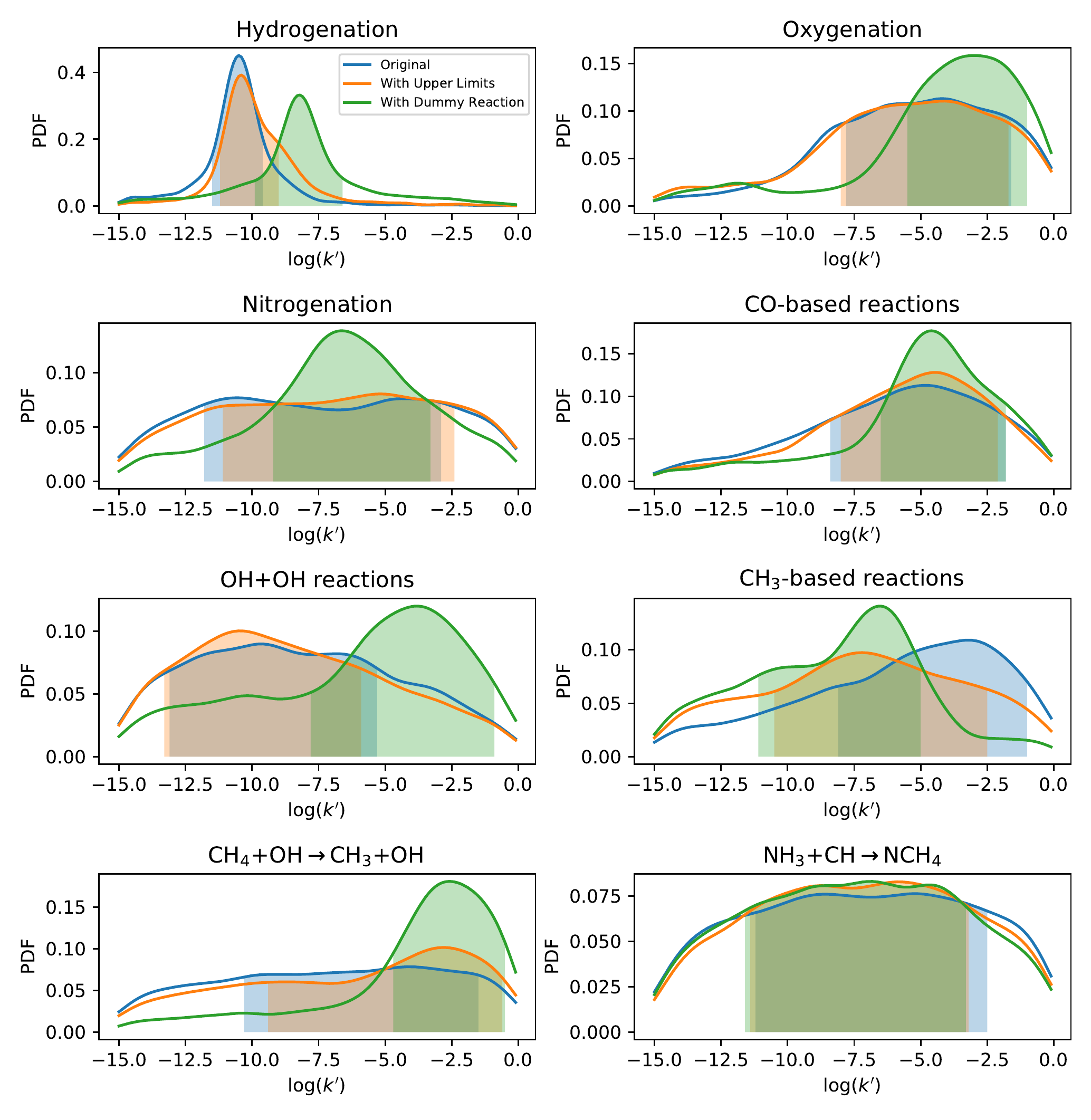}
    \caption{Marginalised posterior distributions for the first eight reaction rate parameters.}
    \label{k_densities_1}
\end{figure*}

\begin{figure*}[h]
    \centering
    \includegraphics[width=\textwidth]{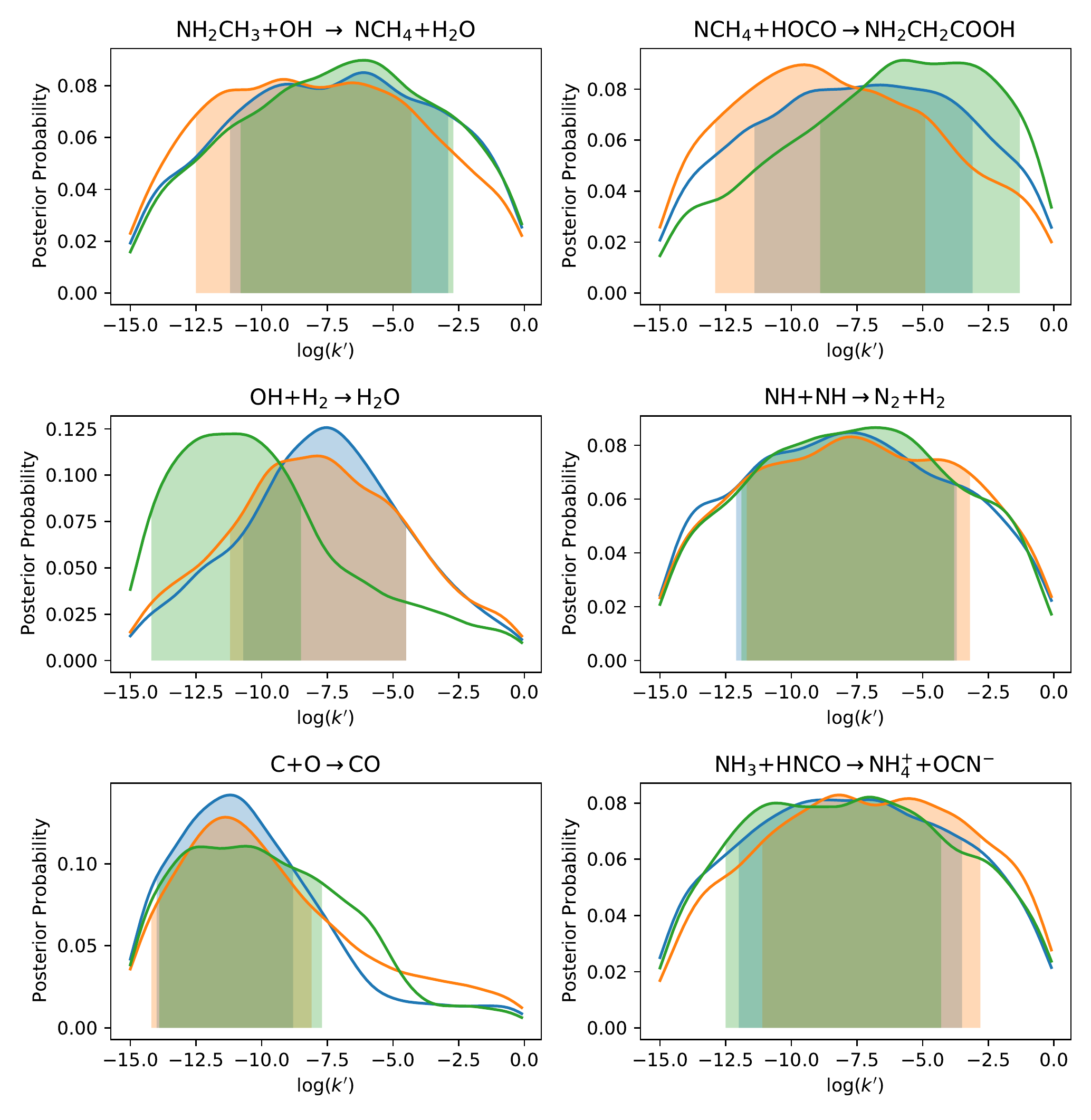}
    \caption{Marginalised posterior distributions for the remaining six reaction rate parameters.}
    \label{k_densities_2}
\end{figure*}

We find that all 14 parameter distributions are non-uniform. As such, this means that we have gained information about the entirety of our 49-D reaction network. By exploiting our knowledge of the grain-surface diffusion mechanism and assuming that reaction rates are dominated by the diffusion rates of a subset of molecules, we have been able to significantly reduce the dimensionality of our problem, therefore making it computationally tractable for the sampler. Figures \ref{k_densities_1} and \ref{k_densities_2} show the marginalised posterior distributions for the reaction rates with the 65\% HDR being the shaded regions when we use the likelihoods expressed in Equations \ref{likelihood} and \ref{upper_limit_likelihood}. For all of the posteriors, the 65\% HDR lies away from the boundaries of the uniform distribution, implying that our choice of prior was appropriate.  We choose not to consider 2-D marginalised posterior distributions, due to the fact that the parameters correspond to groups of reactions as opposed to individual reactions. We consider the frequentist properties of the estimators in Appendix \ref{frequentist_properties_bayesian_estimator}.

There are some noticeable differences in the marginalised posterior distributions when the upper limits are included. The fact that the oxygenation and nitrogenation reaction rate distributions do not significantly change with the inclusion of the upper limits on O$_{2}$ and N$_{2}$ is surprising. One would expect that the reactions \ce{O + O -> O_{2}} and \ce{N + N -> N_{2}} would be the dominant formation mechanisms. As such, it is possible that the upper limits on the abundances of these species may not be constraining enough to affect the obtained posterior distributions. In Appendix \ref{binding_energy_as_function_of_constraint} we explore the distribution of the maximum-posterior binding energy as we vary the weak constraints for the aforementioned four species with upper limits. We also consider how the relative uncertainty on these four abundance measurements affects the obtained values.

We observe that the posterior for the reaction rate of hydrogenation is the most constrained in that it rules out more of the prior parameter space than any of the other posteriors do. In \cite{Heyl}, the lower uncertainty on hydrogen's posterior was related to the size of the constraints on the species formed by hydrogenation, in particular the constraint on water, which is known to have an abundance greater than 0 at the $3.1\sigma$ level. It would make sense that this low level of uncertainty on the constraint drives the low uncertainty on the hydrogenation reaction rate posterior, as it penalises the likelihood function more. In the limit of the uncertainties on the molecular abundances going to zero, one would expect the the posterior distribution of the relevant reaction rate to look like a Dirac delta function. 

\subsection{Binding Energy Posteriors}
\begin{figure*}[h]
    \centering
    \includegraphics[width=\textwidth]{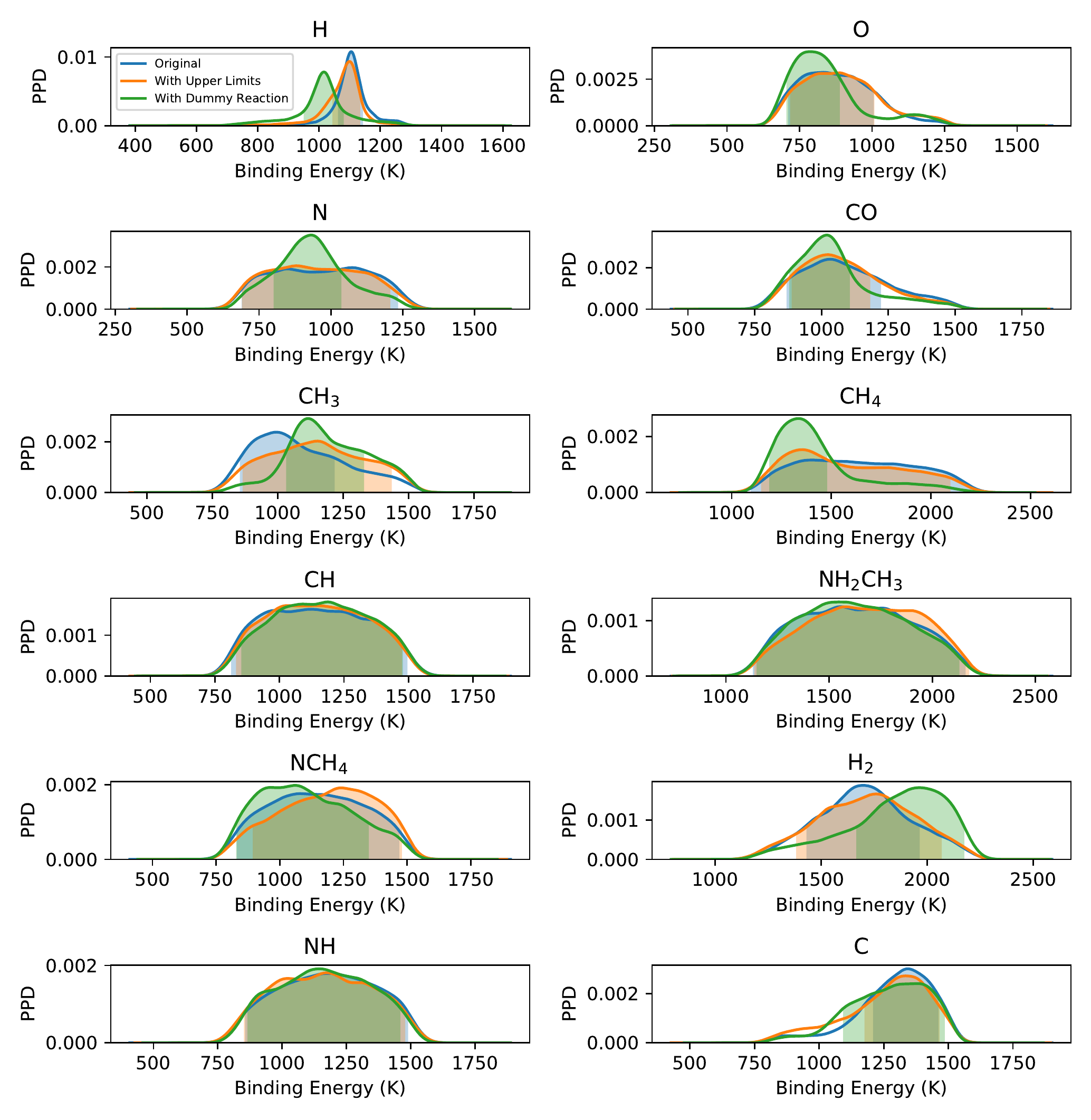}
    \caption{Marginalised posterior probability distributions (PPDs) for the binding energies of the species of interest. The marginalised posterior distributions are also plotted for the case where a dummy reaction for hydrogen is included in the network.}
    \label{binding_energies}
\end{figure*}

The advantage of inferring the reaction rates as opposed to directly inferring the species binding energies is that the reaction rate posteriors make no assumption about the exact nature of the reaction mechanism. One can then select specific reactions which one believes occur via diffusion, thereby reducing the dimensionality of the problem. The list of species that were thought to diffuse are listed in Table \ref{binding_energy_table}. These are calculated from the reaction rate posteriors by solving equation \ref{equation_to_solve_for_E_D}, with the posteriors shown in Figure \ref{binding_energies}. In Table \ref{binding_energy_table}, these binding energies are compared to the values used in \cite{UMIST}, \cite{Penteado} and \cite{Wakelam}. We make use of the posteriors obtained using Equation \ref{upper_limit_likelihood}. This first round of inference is referred to ``Binding Energy 1". We observe that there are no significant differences in the binding energy distributions for most species when we include the upper limits, with the except of CH$_{3}$, for which we see that the inclusion of the upper limits results in a significantly decreased estimated binding energy.

For most of the species for which there are literature binding energies, there is agreement with at least one literature value and the uncertainty on the values is lower than the spread of literature values. No values for the binding energies of NCH$_{4}$ and NH$_{2}$CH$_{3}$ were found in the literature. The binding energies for O and N were both found to be lower than that of H. This is surprising as the reactions of the species with H were classified as hydrogenations in Table \ref{table_of_groupings}. 

However, the binding energies of $H$ and $H_{2}$ were found to differ greatly from the literature binding energies. For the latter, this is related to the fact that there is only a single reaction that $H_{2}$ is consumed in: \ce{OH + H_{2} -> H_{2}O}. The production of water is likely to be dominated by hydrogenation, due to the fact that $H$ is so much more abundant. Furthermore, for this reaction $H_{2}$ must compete with many other molecules to react with $OH$. As such, the amount of water produced through this pathway is less than the amount produced through hydrogenation, which means its reaction rate will be lower than it should be. This results in the high binding energy. 

For some of the species, there is a large variance in the posteriors. This can be attributed to the lack of enough constraints in the network. To demonstrate this, the upper limits on the species N$_2$, O$_2$, H$_2$O$_2$ and glycine were replaced with weak constraints that were derived by halving the upper limit with a 50\% relative uncertainty. It was found that the uncertainties for most species substantially decreased. This could be attributed to the fact that most of the constrained species were formed through hydrogenation, hence why hydrogen's binding energy is so much more well-constrained. This appeared to suggest that the inclusion of these constraints of species not formed solely through hydrogenation would help reduce the variance.

It should be noted that even amongst the literature values, there is not always agreement on the values of the binding energies. The tension in the values can be attributed to varying assumptions made about the grains, such as the ice composition. Additionally, recent work by \cite{Bovolenta} and \cite{binding_energy_distribution} suggests that it might be more appropriate to consider binding energy distributions that vary as functions of the individual binding site. In our work, we have assumed that there is a single binding energy value, which implies that the grains are uniform in nature. In reality, this is unlikely to be true and will need to be accounted for in future work.

\section{The Binding Energy of Hydrogen}
We observe that, despite the high precision of the hydrogenation rate estimate,  the binding energy of hydrogen is inaccurate and does not match any of the literature values within the error. We now look to address this. 

The rate equation approach does not consider positional dependence of species, i.e. it assumes everything can react with everything else on the grain. This might be problematic for H, as there is so much of it, but only a small amount is on the grain mantle. This will not be considered here, as it is outside the scope of the work. 

A rigorous solution would be to account for the formation and subsequent chemical desorption of H$_{2}$. This would take H out of the system and might be more physically realistic. Most of the products of hydrogenation in this network are species for which we have abundances. This means that in order to satisfy all these constraints, the hydrogenation rate posterior will be far too well-constrained. Note that this reaction network is not complete. We can choose to add a “dummy reaction” of the form \ce{H + X -> HX} to represent all the possible reactions involving hydrogen. Notice that these will not necessarily all involve hydrogenation of grain species, but will also include desorption of the produced species. This is why the dummy reaction is not assumed to have the same reaction rate as all the hydrogenations. By leaving the reaction rate of the dummy reaction as an additional free parameter, we can increase the variance of the posterior distribution of hydrogenation and therefore its binding energy posterior.

\subsection{Including Chemical Desorption of $H_{2}$}
The energy released in the reaction of \ce{H + H -> H_{2}} can cause the product to desorb into the gas phase. An estimate for the fraction of $H_{2}$ released was determined in \cite{Minissale} to be: 

\begin{equation}
\eta_{CD} = \exp\left(-\frac{E_{D}N_{dof}}{\epsilon_{CD}\Delta H_{R}}\right), 
\end{equation}
where $E_{D}$ is the desorption energy of the reacting species $\Delta H_{R}$ is the enthalpy of the reaction, $N_{dof} = 3 \times n_{atoms}$ and $\epsilon_{CD}$ is the fraction of kinetic energy the product has as it bounces off the grain surface to escape the potential well. The latter is defined as: 

\begin{equation}
\epsilon_{CD} = \frac{(m-M)^{2}}{(m+M)^{2}}, 
\end{equation}

where $m$ is the mass of the product and $M$ is the effective mass of the grain surface, which is taken to be 120 amu in this work. 

For the chemical desorption of $H_{2}$, $\eta_{CD}$ was found to be roughly 0.9 and this additional loss term due to desorption was included in the differential equation for $H_{2}$. However, it was found to not have a significant impact on the reaction rate and binding energy posteriors. This was a surprising result, but was attributed to the fact that there is far more $H$ in the system than any other species, including H$_{2}$. 

It is also possible that $H_{2}$ formation via this reaction is dominated by the Eley-Rideal mechanism, in which a gas-phase molecule reacts with a grain-surface species \citep{non_diffusive_mechanisms, ER_paper}. This would indicate a weakness of the computational model used which decouples the gas and grain chemistries. While this was done in order to significantly reduce computational runtime and therefore significantly reduce the runtime for the Bayesian inference, highly abundant species such as $H$ and $H_{2}$ are likely to not be accurately described by a decoupled model as they are likely to move between these two phases quite a bit.

\subsection{Including a Dummy Reaction in the Network}
We consider the effect of including the dummy reaction \ce{H + X -> HX} on the entire network. The posteriors for the reaction rates are shown in Figures \ref{k_densities_1} and \ref{k_densities_2} with the corresponding binding energy posteriors being shown in Figure \ref{binding_energies} and listed in Table \ref{binding_energy_table} as ``Binding Energy 2".  

Hydrogen is a unique species, as it is so much more abundant than any other species. Combining this with its higher mobility means that it can react with a wide range of species on the grain. As such, it is important to try and accurately model its behaviour on the grain by accounting for all the possible reactions it can participate in. This dummy reaction acts as a sink for all the excess reactions by accounting for all the other reactions it can participate in. 

There are some differences between the previous posteriors and the ones produced using the dummy reaction. As expected, the hydrogenation reaction rate's posterior sees an increase in its variance. This then translates to an increase in the estimated variance of hydrogen's binding energy. This is not unexpected. Since both $X$ and $HX$ are unconstrained, the amount of hydrogen that is consumed by this reaction is also unconstrained. Therefore, this places a significant uncertainty on the amount of hydrogen that is available for other reactions, thereby inflating the uncertainty. However, even with this increased variance, the binding energy from \cite{Penteado} is not matched within the error. For many of the other parameters, we observe a decrease in the variance of the posteriors through the inclusion of the dummy reaction, with CH$_{4}$ seeing its HDR size shrink significantly through this hydrogen sink. A similar observation can be made for the binding energy posteriors of O, N, CH$_{3}$ and CO.

\section{Application to a Gas-Grain Chemical Code}\label{UCLCHEM_full}

\begin{figure*}[h]
    \centering
    \includegraphics[width=\textwidth]{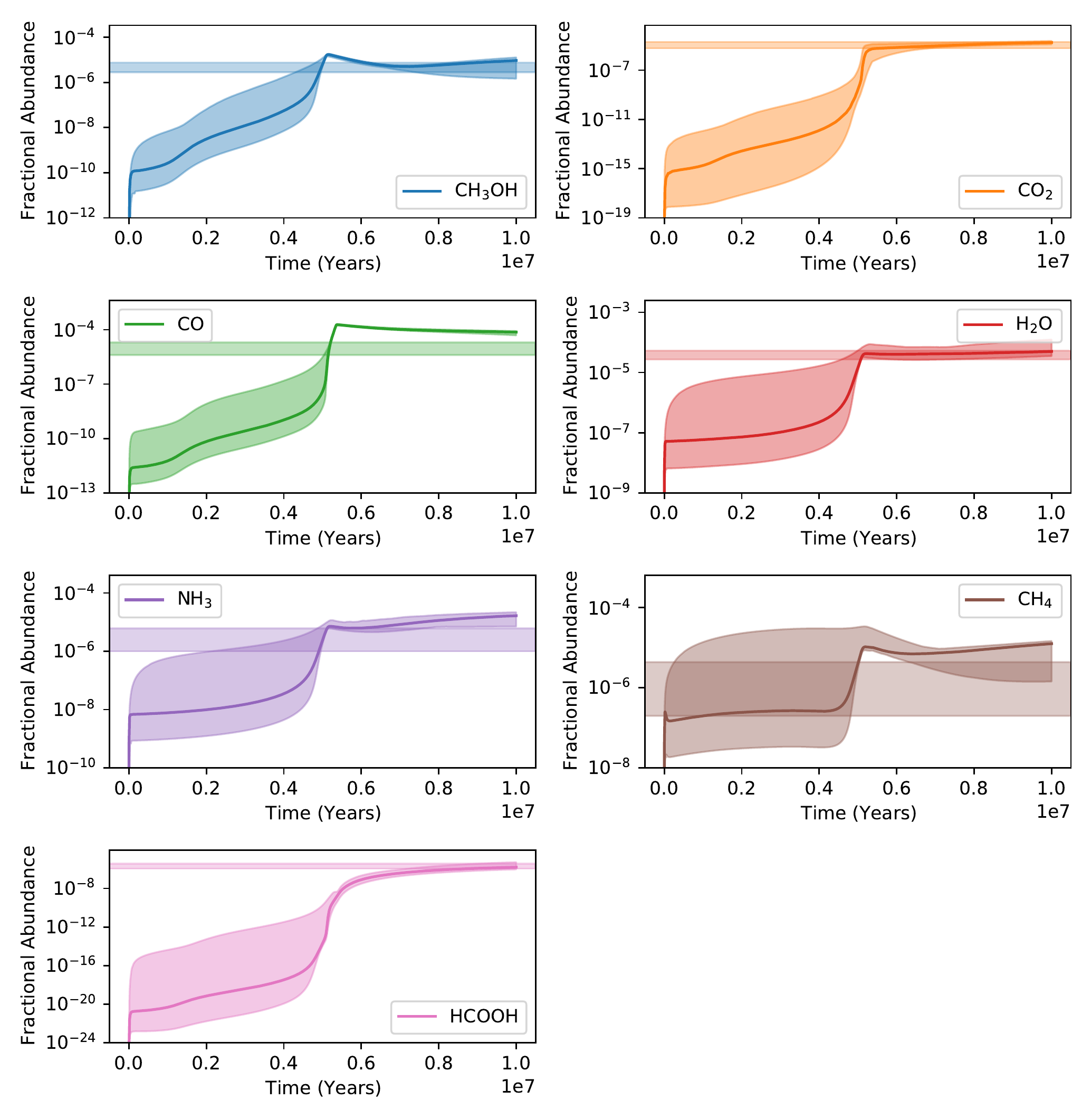}
    \caption{Time series of the fractional abundances for H$_{2}$O, CO, CO$_{2}$ and CH$_{3}$OH. The binding energies for each species were sampled from the marginalised posterior distributions and inputted into the full UCLCHEM code. The horizontal shaded regions are the corresponding measured molecular abundances with their 67\% confidence interval. The time series are plotted with their 95\% confidence intervals. }
    \label{time_series}
\end{figure*}

In this section, we will look to use the binding energies obtained in this work in a full version of the gas-grain chemical code UCLCHEM. This is a form of model-checking. To do this, we sample from the binding energy posteriors in Figure \ref{binding_energies} and input these into UCLCHEM. We aim to determine how well the abundances of species of interest are recovered when the binding energies obtained in this work are inputted into the full gas-grain version of UCLCHEM. Figure \ref{time_series} shows the time series evolution of the fractional abundances of H$_{2}$O, CO, CO$_{2}$, CH$_{3}$OH, NH$_{3}$, CH$_{4}$ and HCOOH, with the 95\% confidence interval for the time series also shown. These are species from Table \ref{abundance_table} that have observed abundances, not simply upper limits that have been converted into weak measurements. The only species with an observed value that has not been included is NH$_{4}^{+}$, but this is not expected to form via diffusion. The purpose of this section is to see how well the inferred binding energy values help in recovering the abundances in a general gas-grain network. We observe that at late times, the final abundances of most of the species become less affected by the binding energies than for earlier times, suggesting that we approach an equilibrium point at a temperature of 10K. If we were to consider a warmer core, it is likely that our final abundances would be different.

The final abundances of H$_{2}$O, CO$_{2}$, CH$_{3}$OH, CH$_{4}$ and HCOOH match the measured values, within the 1$\sigma$ error. This is not the case for NH$_{3}$ and CO. The final abundance for NH$_{3}$ is $1.2\sigma$ from the mean, which is a relatively small discrepancy. On the other hand, CO's final abundance is $4.6\sigma$ from the mean. One possible reason for this is that all the other molecules which are constrained are stable molecules that are unlikely to be depleted sufficiently at 10 K. In contrast, CO is a radical and is likely to react with other radicals. The fact that CO appears to be overproduced here suggests that the network being employed is incomplete. A more complete network would have more CO-based reactions that would lower the CO abundance. Despite, an incomplete network, uncertain elemental abundances or the gas-grain decoupling all contributing to this systematic error, the final abundance for CO is still a sensible value. Overall, knowledge of the diffusion mechanism has allowed us to not only reduce the dimensionality by grouping reactions, but also recover the observed values more precisely compared to \cite{holdship}, where each reaction rate was inferred separately.  

However, despite the discrepancy with CO, the binding energies obtained using the decoupled code provide reasonable results when input into the full gas-grain chemical code. The next step would be to infer the binding energies directly from the full gas-grain code, though this is complicated by the fact that each evaluation of UCLCHEM, which models the full evolution of the cloud, takes of the order of a minute, suggesting a statistical emulator might need to be used to perform the inference in a reasonable amount of time. 

\section{Conclusions}
In this work, we used the diffusion mechanism formalism to significantly reduce the dimensionality of the inference problem, reducing the number of reaction rates to be estimated from 49 to 14. A statistical emulator was trained to further reduce the time taken per forward model evaluation. It was found that the reaction rate of many reactions is ultimately driven by the hopping rate of the more mobile species, thereby allowing us to group several reactions into classes. In doing so, the reaction rate posteriors obtained could be converted into binding energy posteriors for the corresponding mobile species driving the reaction.

This approach yielded binding energy values that were consistent with literature values. The notable exceptions were the binding energies of $H$ and $H_{2}$, whose binding energy values were found to be significantly higher than other literature values. This discrepancy was attributed to issues relating to the chemical model used, which decoupled the gas and grain chemistries in the interest of reducing the time taken for the evaluation of the forward model and therefore the time taken for the inference process. While chemical desorption was found to not have a significant effect on the discrepancy, using a dummy reaction of the form $\ce{H + X -> HX}$ to account for all the possible other reactions involving $H$ somewhat reduced the discrepancy, but not enough. 

This work has developed an important step in estimating reaction rate parameters using Bayesian inference. It was seen that dimensionality will scale slower than the number of reactions. This reduces the number of samples that are needed to reach a stationary posterior. This approach can be trivially expanded to include more complex reaction networks. This will prove particularly important in the context of considering the formation chemistry of glycine or other amino acids. The formation routes are likely to contain a large number of diffusion reactions. However, inferring the reaction rates will not become unfeasible, due to how the dimensionality scales with the number of reactions.

In Section \ref{UCLCHEM_full}, we sampled from the obtained binding energy posteriors and input these binding energies into the full gas-grain version of UCLCHEM. We found some agreement between the obtained molecular abundances and the observed values. However, if one wished to infer from UCLCHEM directly, one would need to account for the fact that the inference process would take longer, on account of one evaluation of the full version of UCLCHEM taking of the order of a minute compared to the 0.5 seconds that is typical of the simplified code used in this work. Future work will look to employ statistical emulation to the full version of UCLCHEM to circumvent this problem. Alternative sampling techniques that are adaptive could be utilised to this work's emulator \citep{adaptivesampling}.

Further work will need to consider larger grain-surface networks and include gas chemistry. Additionally, one should look to consider other non-diffusive grain-surface reaction mechanisms. Expressions for these reaction rates have been formulated in \cite{non_diffusive_mechanisms}. Including these would ensure that a more accurate picture of the chemistry would be obtained. It would also be interesting to investigate the validity of the claim that the measurements are Gaussian-distributed, as this has a direct impact on the formulation of the likelihood function. This assumption would have an impact on the posteriors obtained. There exist various methods to perform Bayesian inference without requiring the specification of a likelihood function, such as Approximate Bayesian Computation that are the focus of current work.

Further work would also need to address the lack of sufficient abundance data. It is clear that the abundances of more species need to be known in order to better constrain the reaction rate posteriors as well as the binding energy posteriors.



\acknowledgments
The authors thank the referees for their constructive
comments that greatly improved this work. J. Heyl is funded by an STFC studentship in Data-Intensive Science (grant number ST/P006736/1). This work was also supported by European Research Council (ERC) Advanced Grant MOPPEX 833460. S. Viti acknowledges support from the European Union’s Horizon 2020 research
and innovation programme under the Marie Skłodowska-Curie grant
agreement No 811312 for the project ``Astro-Chemical Origins” (ACO). This work used computing equipment funded by the Research Capital Investment Fund (RCIF) provided by UKRI, and partially funded by the UCL Cosmoparticle Initiative.
\software{UCLCHEM \citep{UCLCHEM}, PyMultiNest \citep{pymultinest}, F2PY \citep{F2Py}} 



\clearpage
\appendix
\section{Evaluating the Frequentist Properties of the Bayesian Estimators}\label{frequentist_properties_bayesian_estimator}
\begin{figure*}[h]
    \centering
    \includegraphics[width=\textwidth]{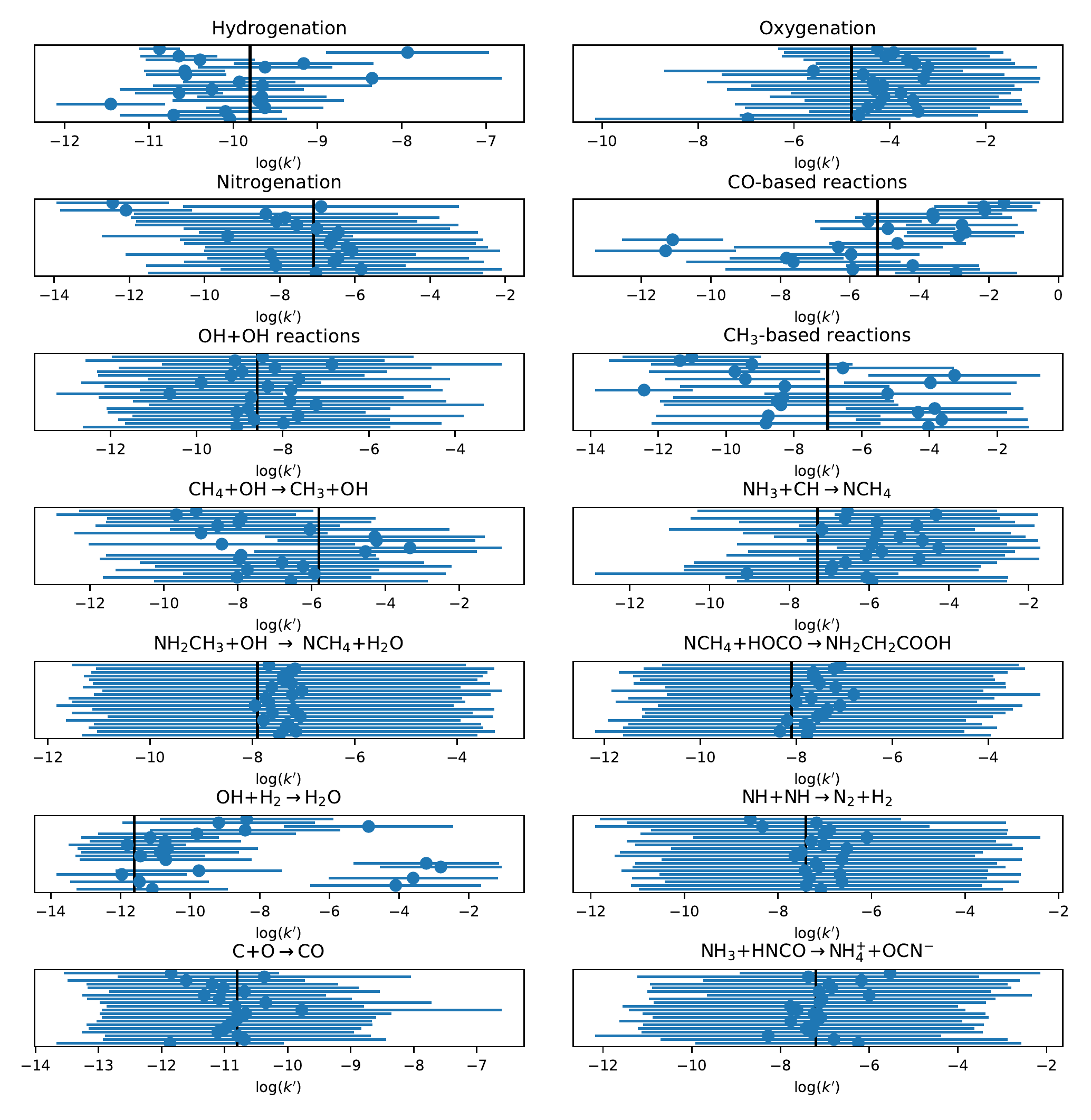}
    \caption{A strip plot of the how well the reaction rates are recovered when the forward model is run with some noise on the reaction rates. The vertical black line in each plot represents the ``true" reaction rate.}
    \label{frequentist_plot}
\end{figure*}

In this section, we seek to determine whether the constraints imposed are significantly influencing the resulting posterior distribution. To determine this, we run the forward model using reaction rates drawn from a Gaussian distribution with a mean value equal to the maximum-posterior reaction rate obtained in this work (which represents the ``true" reaction rate) and a standard deviation equal to 1. Bayesian inference was then used to recover these reaction rates. This was repeated 20 times using the statistical emulator. The strip plots in Figure \ref{frequentist_plot} show the values of the reaction rates recovered with the associated uncertainties. This analysis is meant to demonstrate to what extent the constraints imposed by Equation \ref{upper_limit_likelihood} are influencing the posteriors obtained.

It becomes clear that the extent to which the constraints affect the posteriors depends on the parameter. For the hydrogenation reaction rate, we see that the 65\% highest density regions contain the true reaction rates used in the simulation 65\% of the time, that is for 13 of the 20 strips. We find that the high density regions are jittered around the true value, suggesting there is no bias. Overall, what we find is that the constraints are significantly influencing the hydrogenation reaction rate posterior. This is perhaps unsurprising given that most of the constrained species are products of hydrogenation. It is also for this reason that the binding energy for hydrogen has the lowest uncertainty.

However, when the other posteriors are considered, it is clear that there is a greater level of prior domination. While the HDRs are all significantly smaller than the prior range from -15 to 0, we still find that within the HDR the posteriors are not as sharp as for the hydrogenation. This can be confirmed visually by considering the posteriors shown in Figures \ref{k_densities_1} and \ref{k_densities_2}. While some of the strips, such as for CO-based reaction, show more jitter around the true value, it is clear that more data is required to counter the influence of the prior distribution. However, we observe that there is no bias in the obtained posteriors.

\section{Determining the Effect of Constraints on the Inferred Binding Energy Values}\label{binding_energy_as_function_of_constraint}
\begin{figure*}[h]
    \centering
    \includegraphics[width=\textwidth]{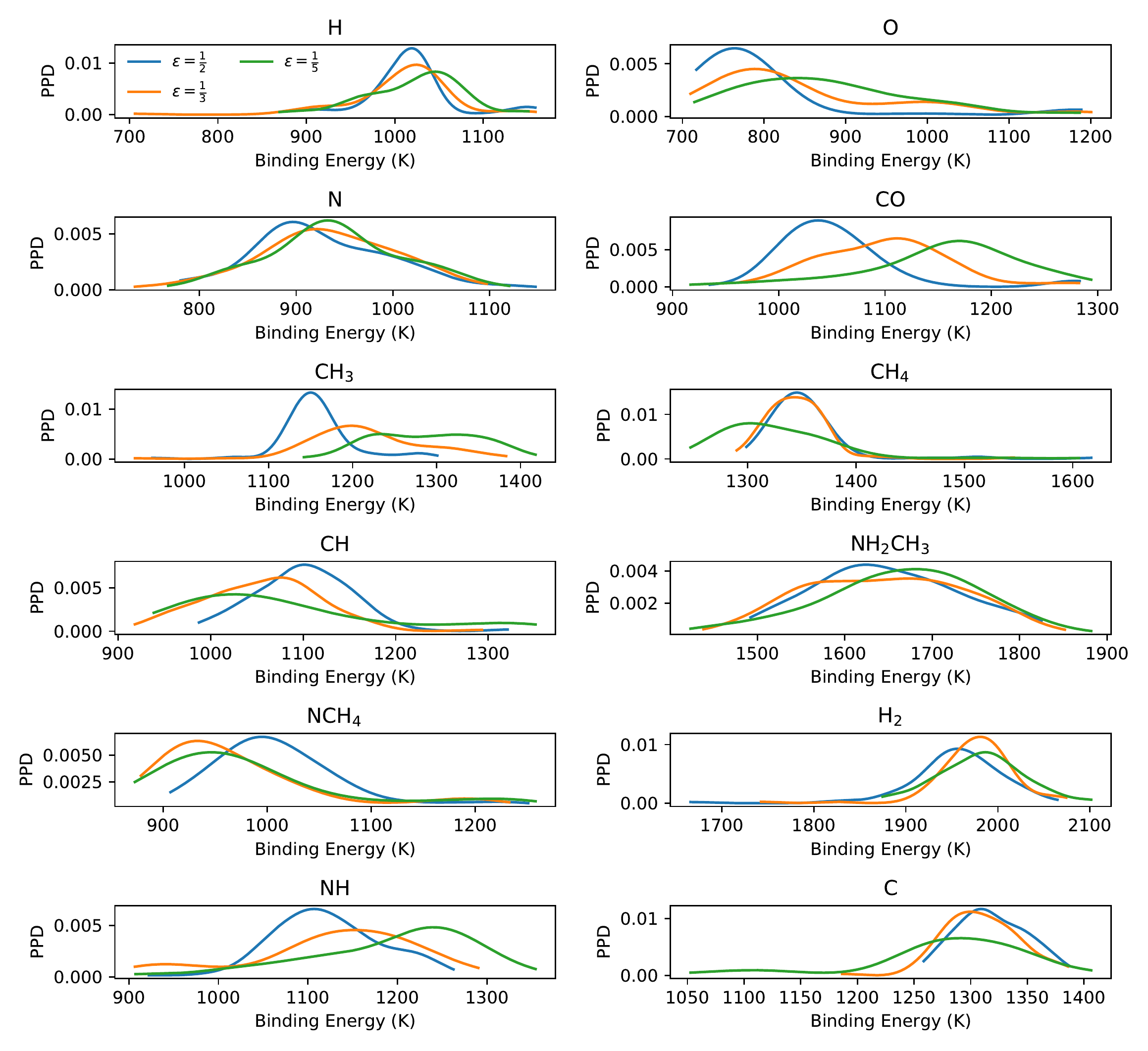}
    \caption{Distributions of the maximum-posterior binding energies obtained when the constraints on N$_2$, O$_2$, H$_2$O$_2$ and glycine are varied.}
    \label{binding_energy_distrbutions_different_constraints}
\end{figure*}

We observed that the inclusion of upper limits in the likelihood function given by Equation \ref{upper_limit_likelihood} did not have a significant bearing on the binding energy posterior distributions. This suggests that the upper limits listed in Table \ref{abundance_table} for N$_{2}$, O$_{2}$, H$_{2}$O$_{2}$ and glycine may not be sufficiently constraining. In that case it might be more useful to have abundance measurements for these species. To test the effect of these abundance values on the obtained binding energies, we ran the inference 1000 times using the statistical emulator and plotted the distribution of the maximum-posterior binding energy values in Figure \ref{binding_energy_distrbutions_different_constraints}. For each inference run, the constraints for each of the species with upper limits in Table \ref{abundance_table} was taken to be a random value between 0 and the upper limit. These abundance values were sampled uniformly in this range. The relative error on these four measurements was varied to equal 50\%, 33.3\% and 20\%. The relative errors are represented as $\epsilon$.

We observe that the size of the relative errors has some bearing on the maximum-posterior binding energy values obtained as well as the spread of values. For most of the species we observe that there is a significant increase in the spread of inferred values. This demonstrates the importance of detecting further grain-surface species in order to better constrain the binding energy values.

\bibliography{references}{}
\bibliographystyle{aasjournal}

\end{document}